\documentclass{elsart}

\usepackage{graphicx}
\usepackage{amssymb}

\begin{document}

\begin{frontmatter}

\title{Chemical Bonding and Charge Distribution at Metallic Nanocontacts}

\author{U.~Schwingenschl\"ogl\corauthref{cor1}},
\ead{Udo.Schwingenschloegl@physik.uni-augsburg.de}
\author{C.~Schuster}
\corauth[cor1]{Corresponding author. Fax: 49-821-598-3262}
\address{Institut f\"ur Physik, Universit\"at Augsburg, 86135 Augsburg,
Germany}

\begin{abstract}
We present results of electronic structure calculations for aluminium
contacts of atomic size, based on density functional theory and the local density
approximation. Addressing the atomic orbitals at the neck of the
nanocontact, we find that the local band structure deviates strongly from
bulk fcc aluminium. In particular, hybridization between Al $3s$ and $3p$
states is fully suppressed due to directed bonds at the contact. Moreover,
a charge transfer of 0.6 electrons off the contact aluminium site is found.
Both the suppressed hybridization and the violated charge neutrality are
characteristic features of metallic nanocontacts. This fact has serious
consequences for models aiming at a microscopic description of transport
properties.
\end{abstract}

\end{frontmatter}

\section{Introduction}

Metallic nanocontacts nowadays are prepared by means of scanning tunneling
microscopy \cite{pascual93} or break junction techniques \cite{scheer97}.
In both cases, a wire is streched with a precision of a few picometers by
means of piezoelectric elements until finally a single-atom configuration
is reached. Such contacts have attracted great
attention over the last couple of years, in particular as concerns their
electrical transport properties. Since transport is restricted to a small
number of atomic orbitals at the contact, conductance across metallic
nanocontacts strongly depends on the local electronic structure. An
atomic-sized constriction between two electrodes can accomodate only a
small number of conducting channels, which is determined by the number of
valence orbitals of the contact atom. In addition, the transmission of
each channel is fixed by the local atomic environment.

A detailed review on the quantum properties of atomic-sized conductors is
given by Agra\"it {\it et al.} \cite{agrait03}, who report on a
large number of investigations of the electronic structure
of metallic nanocontacts. From the theoretical point of view, the
conductance of nanocontacts can be analyzed by means of band structure
calculations. However, several theoretical studies reported in the
literature are based
on less accurate approaches for calculating the electronic structure, as
the tight-binding model, for instance \cite{cuevas98a,cuevas98b,pauly06}.
Analyses of transmission eigenvalues or conducting channels based on such
methods are expected to be of limited validity.

Using state-of-the-art electronic structure calculations, we show in this
letter that spacial restrictions of the electronic states at the
nanocontact have serious consequences for the local band structure. In
particular, for aluminium nanocontacts we find that hybridization between
Al $3s$ and $3p$ states -- characteristic of the bulk material --
disappears completely due to directed bonds at the neck of the contact.
Furthermore, the inhomogeneity of the contact geometry results in large
deviations from the charge neutrality expected for metallic compounds.

\section{Calculational Details}

The electronic structure calculations presented in the following are
based on density functional
theory and the local density approximation. To be specific, we use the
augmented spherical wave method \cite{eyert00}, which has shown to be
particularly suitable for studying effects of covalent bonding and
hybridization \cite{us03,eyert04,schmitt05,eyert05}. The spherical wave
basis set allows intuitive interpretation of the band structure results in
terms of atomic orbitals. In order to model the correct shape of the
crystal potential in the voids of the nanocontact, additional augmentation
spheres are included. The final basis set thus comprises Al 3$s$, 3$p$, and
3$d$ states, as well as states of the additional spheres. For the Brillouin
zone sampling, we use an increasing number of {\bf k}-points within the
irreducible wedge, ranging from 8 to 144, to ensure convergence of the
results with respect to the {\bf k}-space grid.

\begin{figure}
\centering 
\includegraphics*[width=0.5\textwidth]{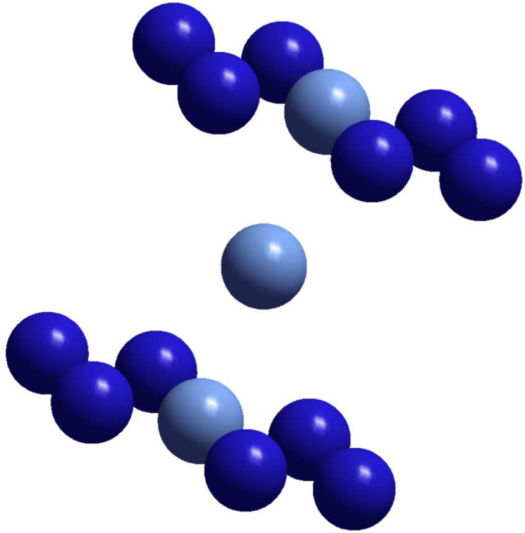}
\caption{Atomic configuration of the aluminium nanocontact used as input
for the LDA electronic structure calculations. The contact site and its
next-nearest neighbour sites are highlighted.}
\label{fig1}
\end{figure}

Our calculations rely on the contact geometry depicted in figure \ref{fig1}.
The central Al site is connected to planar Al units on both sides, each
consisting of seven atoms in a hexagonal arrangement (fcc [111] orientation).
The central sites of these planar units lie on top
of the contact atom, thus forming directed bonds along the
crystallographic $c$-axis. Finally, the finite Al units are connected to
Al $ab$-planes of infinite extension, which enables
us to apply periodic boundary conditions.

A convenient choice for setting up the details of the contact geometry
is the use of bulk (fcc) aluminium bond lengths and bond angles, on which
the subsequent results and discussion of the electronic structure are based.
To study structural relaxation effects, we additionally have minimized
the atomic forces in our contact using the WIEN$2k$ program package, a
full-potential linearized augmented-plane wave code \cite{wien2k}. Due to
the symmetry of the contact geometry, the structural optimization does
not change the 180$^\circ$ bond angle between the contact Al site and its
next-nearest neighbours. Importantly, the related Al-Al bond lengths at the
neck of the contact increase only slightly by less than $0.01$\,\AA, which
hardly affects the electronic states.

\section{Electronic Structure Results}

For bulk aluminium it is well established that the formal Al 3$s^2$3$p^1$
electronic configuration is seriously
interfered by hybridization effects, giving rise to a prototypical
$sp$-hybrid system. In particular, this fact is reflected by the occupation
of the valence states as resulting from LDA electronic structure
calculations \cite{eyert97}. Both the 3$s$ and the three-fold degenerate
3$p$ states cover the same energy region with respect to the Fermi level.
As a consequence, the valence electrons are almost equally distributed
between these orbitals.

\begin{figure}
\centering 
\includegraphics*[width=0.5\textwidth]{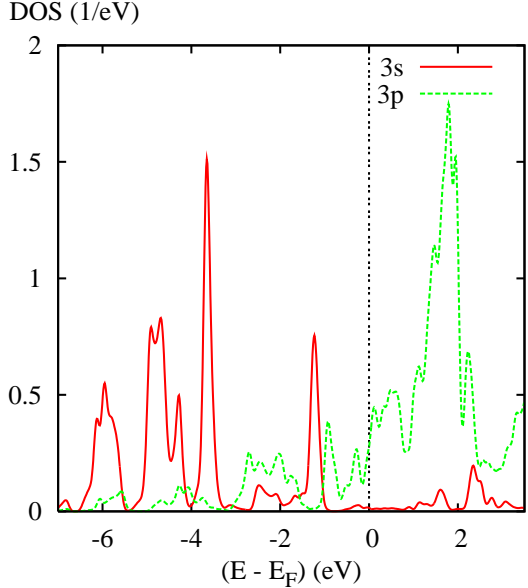}
\caption{Partial Al $3s$ and $3p$ densities of states for the aluminium
site at the neck of the nanocontact. Most of the occupied valence states
are of $3s$ type.}
\label{fig2}
\end{figure}

However, the above situation changes dramatically when covalent bonding
is no longer isotropic but becomes restricted to specific directions, which
happens in the case of an atomic-sized contact. Partial Al $3s$ and $3p$
densities of states for the central Al site of our nanocontact as resulting
from the LDA calculation are shown in figure \ref{fig2}. While most of the
occupied states are of $3s$ type, the $3p$ states dominate at energies
above the Fermi level. To be more specific, about 70\% of the valence
electrons occupy the $3s$ atomic orbital, which strictly contradicts the
equal distribution in the case of bulk aluminium and consequently spoils an
interpretation in terms of $sp$-hybrid states. Since figure \ref{fig2} shows
almost no contributions of $3s$ and $3p$ states at energies dominated by
the other states, respectively, evolution of hybrid orbitals is precluded.

\begin{figure}
\centering 
\includegraphics*[width=0.5\textwidth]{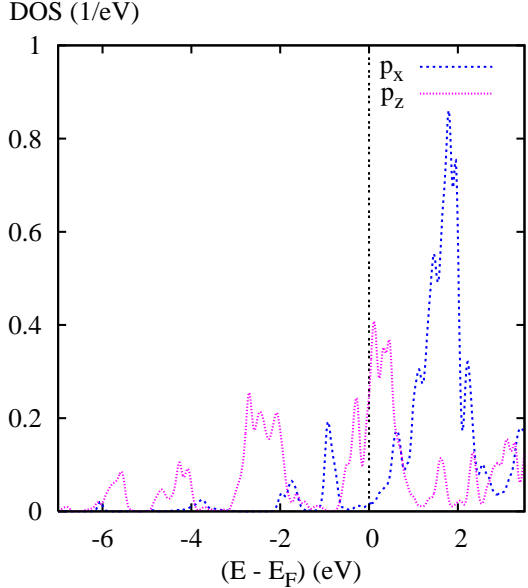}
\caption{Decomposition of the Al $3p$ DOS from figure \ref{fig2} into the
symmetry components $p_x$, $p_y$, and $p_z$. By symmetry, $p_x$ and $p_y$
states are degenerate.}
\label{fig3}
\end{figure}

In order to address the chemical bonding at the neck of the nanocontact
in more detail, figure \ref{fig3} gives a decomposition of the Al $3p$
DOS, see figure \ref{fig2}, into its symmetry components. By symmetry,
$p_x$ and $p_y$ states are degenerate. Most $3p$ electrons occupy the
$p_z$ orbital, which is oriented along the crystallographic $c$-axis and
therefore mediates $\sigma$-type overlap across the contact. Since neither
$s$ nor $p_x$/$p_y$ states give rise to significant contributions to the
DOS at the Fermi energy, chemical bonding at the neck of the nanocontact
is well characterized in terms of directed bonds of $3p_z$ type.

Both the $p_x$ and $p_y$ atomic orbitals at the neck of the nanocontact
do not mediate chemical bonding due to the spacial restriction of the
crystal structure. Their occupation thus is strongly reduced as compared
to the $p_z$ orbital, as well as to the orbital populations found in bulk
aluminium. As a consequence, we cannot expect local charge neutrality at a
nanocontact, despite the fact that the system is metallic. Actually,
our calculation results in a net charge transfer of about 0.6 electrons
off the contact aluminium site.

\section{Discussion and Conclusions}

Our investigation of the electronic properties of aluminium atomic-sized
contacts results in two characteristic features: the suppression of
hybridization and the violation of charge neutrality at the neck of the
contact. Since both features trace back to nothing but the spacial
restriction of the crystal structure due to the contact, they are typical
of metallic nanocontacts in general. However, we note that the amount of
charge transfer off the contact may differ from our aluminium result of 0.6
electrons for other compounds. It likewise may depend on the particular
shape of the nanocontact and therefore on the experimental setup.

The local suppression of hybridization has important implications
for models aiming at a microscopic description of electrical transport
across metallic nanocontacts. It calls for methods accounting for the
details of the contact geometry as concerns local modifications of the
electronic orbitals and their occupations, with respect to the bulk
configuration. In particular, our findings cast serious doubts on
approaches to transmission eigenvalues or conduction channels of
nanocontacts on the basis of tight-binding models
\cite{cuevas98a,cuevas98b,pauly06}.

\section*{Acknowledgements}
We thank U.\ Eckern and P.\ Schwab for fruitful discussions. Financial
support by the Deutsche Forschungsgemeinschaft within SFB 484 is gratefully
acknowledged.

\end{document}